\documentclass[12pt]{article}
\usepackage{epsfig,graphicx}

\title{Effective 4D propagation of a charged scalar particle in Visser brane world\thanks{It is a pleasure to dedicate this work to Alberto Garc\'{\i}a on the occasion of his 60th birthday}}
\author{R. Guzm\'an and H.A. Morales-T\'ecotl\thanks{
Associate member of Abdus Salam ICTP, Trieste, Italy. Email hugo@xanum.uam.mx}\\
Departamento de F\'{\i}sica, Universidad Aut\'onoma Metropolitana
Iztapalapa\\ A.P. 55-534 M\'exico D.F. 09340, M\'exico}

\begin{document}

\maketitle

\abstract 
In this work we extend an analysis due to Visser of the effective propagation of a neutral scalar particle on a brane world scenario which is a particular solution of the five dimensional Einstein-Maxwell equations with cosmological constant having an electric field pointing in the extra spatial dimension. We determine the dispersion relations of  a charged scalar particle to first order in a perturbative analysis around those of the neutral particle. Since  depending on whether the particle is charged or not the dispersion relations change, we could collect bulk information, namely the presence of the electric field, by studying the 4D dynamics of the particles.

\baselineskip=20pt

\section{ Introduction}

A new picture of extra-dimensions emerged recently \cite{reviews} based on the
Horava-Witten eleven-dimensional supergravity \cite{HW}. In this context, the
ordinary matter fields are not supposed to be defined everywhere but, in
contrast, are assumed to be confined in a submanifold, called brane, embedded
in a higher dimensional space. In order to solve the hierarchy problem, it was suggested
by Arkani-Hamed et al \cite{ADD}that we live confined in a three-brane surrounded by $n\geq 2$ (flat and
compact) extra-dimensions with a size $R$ as large as the millimeter scale.

Another proposal, even more interesting from the point of view of general
relativity and cosmology, is due to Randall and Sundrum \cite{RS}. They consider only
one extra-dimension but take into account the self-gravity of the brane
endowed with a tension $\sigma $. The main motivation for exploring
cosmology in models with extra-dimensions is that the new effects might be
significant only at very high energies, e.g. in the very early universe.

Remarkably, Visser discussed an exotic class of five dimensional Kaluza-Klein models \cite{visser} previous to the aforementioned scenarios, in which the internal space is neither compact nor even of
finite volume. Particles are gravitationally trapped near a four-dimensional submanifold of
the higher dimensional spacetime. A particular scenario was obtained in \cite{visser} by solving five dimensional Einstein-Maxwell equations which contain a bulk cosmological constant and an electric field pointing in the fifth direction. Moreover, by giving the Klein-Gordon equation for a neutral scalar particle a Schr\"odinger like form, the problem of obtaining dispersion relations was translated into a  non relativistic  eigenvalue problem involving the Rosen-Morse potential.

In this work we extend the analysis of \cite{visser} to the case of charged scalar particles coupled to the electric field of the scenario. This leads us to four dimensional modified dispersion relations. Since the electric field is  orthogonal to the four dimensional brane this effect can be interpreted as extra dimensional.

The paper is organized as  follows. For the sake of completeness Section 2 gives the details of Visser scenario and  in section 3 the dispersion relations analysis of Visser for the neutral scalar particle is reproduced. Next, the charged scalar particle coupled to the electric field of Visser scenario is considered in Section 4. Finally in section 5 we discuss our results.

\section{Visser brane world}

This scenario is a solution of the five-dimensional Einstein-Maxwell equations with cosmological constant 
\begin{eqnarray}
G_{AB}&=&\Lambda g_{AB}+T_{AB} \label{eq:einstein}\\
 \frac{1}{\sqrt{-g}}\partial _A
\left( \sqrt{-g}F^{AB}\right) &=&0, \label{eq:maxwell}
\end{eqnarray}
where $\Lambda$ is the cosmological constant,
$F_{AB}=\partial_A A_B -\partial_B A_A$ and $T_{AB}$ is the corresponding energy momentum tensor of the electromagnetic field.

Using the following ansatz for the metric and electromagnetic potential \cite{visser}
\begin{eqnarray}
ds_{(5)}^2&=&-{\rm e }^{2\phi (y )}(dt)^{2}+d\vec{x}\cdot d\vec{x}+(dy )^{2}\,, \label{eq:ansatzg} \\
A_{0}&=&a(y ),\quad A_{1}=A_{2}=A_{3}=A_5= 0  \label{eq:ansatzA}
\end{eqnarray}
and taking $y$ as the coordinate of the extra dimension, one gets
\begin{equation}\label{eq:aphi}
{\rm e}^{\phi}=\cosh (Ey )\,,\qquad a=\sinh (Ey ) \,,
\label{eq:ephi}
\end{equation}
where $E$ fulfills $E^2=2 \Lambda$. 

\begin{figure}
\center{\includegraphics[height=90mm,angle=-90]{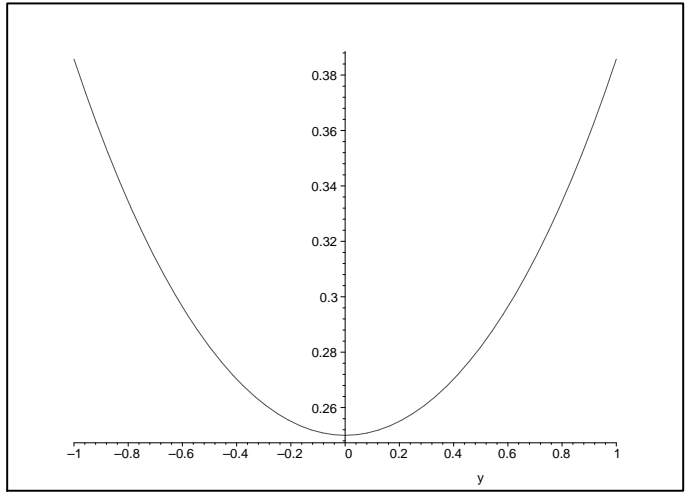}}
\caption{\small The potential $V(y)=\frac{1}{4}cosh(2Ey)$ localizes classical test particles on the brane at $y=0$. Units in the graph are such that $E=1$.} \label{hugofig1}
\end{figure}

Classical test particles follow geodesics corresponding to (\ref{eq:ansatzg}) and (\ref{eq:ephi}) given by
\begin{eqnarray}
 \ddot{t}+2\phi'\dot{y}\dot{t}&=&0, \label{eq:geodesicat}\\
 \ddot{x}^i&=&0, \label{eq:geodesicax}\\
 \ddot{y}+\phi'e^{2\phi}\dot{t}^2&=&0. \label{eq:geodesicay}
 \end{eqnarray}
with $ \cdot = \frac{d}{d\lambda}$. Hence, using $t=\lambda-\lambda_0$, (\ref{eq:geodesicat}) just says $\phi$ is $\lambda$ independent. 
In light of (\ref{eq:geodesicax})-(\ref{eq:geodesicay}), particles move along straight lines on the brane subspace at $y=0$. This is due to the localizing potential in the extra dimension  that appears after expressing (\ref{eq:geodesicay}) in the form (See Fig.\ref{hugofig1})
\begin{eqnarray} 
\ddot{y}=-\frac{d}{dy}V(y)\;,\qquad
V(y)=\frac{1}{4}\cosh(2Ey)\, .
\end{eqnarray}

\section{Neutral scalar particle}
A neutral scalar particle propagating in the Visser scenario described by (\ref{eq:ansatzg}) and (\ref{eq:ephi}), obeys the Klein-Gordon equation
\begin{equation}\label{eq:kgvisser}
[-e^{-2\phi}\partial^2_{tt}+\sum_{i=1}^3\partial^2_{ii}+\phi'\partial^2_{yy}]\Phi=M^2\Phi
\end{equation}
with $M$ being the mass of the particle in five spacetime dimensions. A convenient way of analysis is gained by reducing this equation to an Schr\"odinger like one. Using  
\begin{equation}
\Phi =\exp [-i(\omega -\vec{k}\cdot \vec{x})]{\rm e} ^{-\frac{\phi }{2}}\eta(y )
\end{equation}
yields the eigenvalue equation
\begin{eqnarray}
H\eta&:=&\left[ -\frac{1}{2} \partial_{yy}^{2}+\frac{%
1}{8}E^{2}-\left(\frac{1}{2}\omega ^{2}-\frac{1}{8}E^{2}\right){\rm sech}^{2}(Ey)\right]
\eta =\varepsilon^{(0)} \eta \label{eq:H}\\
\varepsilon^{(0)}&=&-\frac{M^2+\vec{k}^2}{2} \label{eq:varepsilonMk}\,.
\end{eqnarray}
Such a potential is the well known Rosen-Morse and the corresponding eigenvalue problem has been solved. The eigenvalues are
\begin{equation}
\varepsilon ^{(0)}_{n}=-\frac{1}{2}\left\{ \left( \omega -\left[ n+\frac{1}{2}\right]
E\right) ^{2}-\left( \frac{E}{2}\right) ^{2}\right\} \,.
\label{eq:varepsilon}
\end{equation}
The corresponding eigenfunctions turn out to be
\begin{eqnarray}
\eta_n & = & C_n\left( \cosh (Ey) \right) ^{-2\sigma }u_n \\
u_{1n}& = &F(-\sigma +n,-\sigma -n,\frac{1}{2};z)\,,\quad n=0,2,\dots \\
u_{2n}& = &F(-\sigma +n+\frac{1}{2},-\sigma -n+\frac{1}{2},\frac{3}{2};z) \,,\quad n=1,3,\dots \\
z & = & -\sinh^{2}(Ey)
\end{eqnarray}
where $C_n$ are the normalization constants, $F$ represent the hypergeometric functions and $\sigma=\frac{1}{2}\frac{\omega }{E}-\frac{1}{4}$. The normalizability condition requires $F$ to have $n$ an integer. In light of (\ref{eq:H}) there will be localization whenever $\left(\frac{1}{2}\omega ^{2}-\frac{1}{8}E^{2}\right)>0$ namely $\sigma > 0$. Moreover the number of bound states $N$ satisfies $N<2\sigma$, so in order to have at least one it is required that $\sigma>\frac{1}{2}$.

Next it is straightforward to get the dispersion relations. Combining (\ref{eq:varepsilonMk}) and (\ref{eq:varepsilon}) leads to
\begin{equation}
\omega _{n}(\vec{k})=\left[ n+\frac{1}{2}\right] E\pm \left[ m^{2}+\vec{k}^{2}\right] ^{1/2}\,.
\end{equation}
Here $m^{2}=M^{2}+\left( \frac{E}{2}\right) ^{2}$ could be considered as an effective four dimensional mass.

\section{Charged scalar particle}

Now we proceed to extend Visser analisis to the case of a charged scalar particle. To begin with consider the Klein-Gordon equation with coupling to an electromagnetic field in curved spacetime

\begin{equation}
\frac{1}{\sqrt{-g}}(\partial _{A}-ieA_{A})\left[ \sqrt{-g}g^{AB}(\partial
_{B}-ieA_{B})\Phi \right] =M\medskip ^{2}\Phi 
\end{equation}
together with (\ref{eq:ansatzg})-(\ref{eq:aphi}). We obtain 

\begin{eqnarray}\label{eq:kgcargado2}
  [-e^{-2\phi}\partial_{tt}&+&\sum_{i=1}^3\partial_{ii}+\phi'\partial_{yy}+K]\Phi = M^2\Phi \\
K &=&-e^{-2\phi}(2iea\partial_t+e^2a^2)\,.
\end{eqnarray}

Using 
\begin{equation}
\Phi =\exp [-i(\omega t -\vec{k}\cdot \vec{x})]{\rm e} ^{-\frac{\phi }{2}}\chi(y )
\end{equation}
this is equivalent to the eigenvalue equation 
\begin{eqnarray}
(H+H_{1})\chi &=&\varepsilon \,\chi \,,\qquad \qquad \varepsilon =-\frac{%
M^{2}+\vec{k}^{2}}{2} \label{eq:H+H1} \\
H_{1} &=&-e^{-2\phi}(ea\omega+\frac{e^2 a^2}{2})\label{eq:H1} \,.
\end{eqnarray}
with $H$  given by (\ref{eq:H}). 

\begin{figure}
 \centerline{
 {\includegraphics[height=60mm]{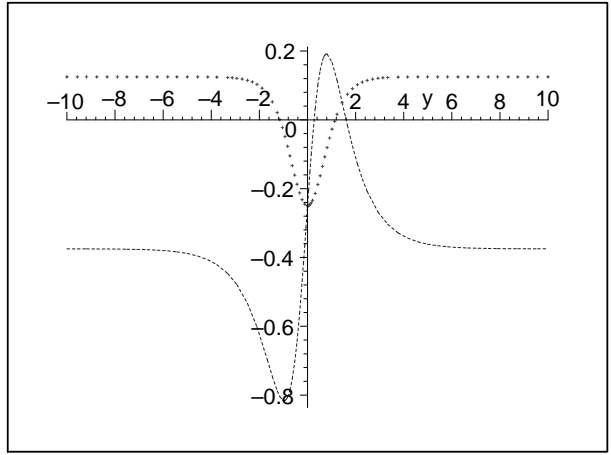}}}
  \caption{{\small The dashed curve represents the Rosen-Morse potential appearing in the neutral scalar particle propagation. The continuous  curve is the potential for the charged scalar particle. Tunneling into the extra dimension can  not occur. We set $E=1, e=1, M=1, \omega=1$.}} \label{metaestados}
\end{figure}

Notice the potential for the charged scalar particle is that of Rosen-Morse (\ref{eq:H}) added to (\ref{eq:H1}). Fig.\ref{metaestados} reveals there can be bound states, namely localized particles on the brane at $y=0$, but they can not tunnel into the extra dimension independently on whether they are neutral or charged.

Lacking an exact solution study of (\ref{eq:H+H1}) we face the problem of finding dispersion relations of the charged particle by considering $H_1$ as a perturbation. Thus we have, to first order,
\begin{eqnarray}
\varepsilon_n &=&\varepsilon_n^{(0)} +\delta \varepsilon_n (e,E,\omega ) \label{eq:QvarepsOvareps} \\
\delta \varepsilon_n (e,E,\omega )&=&\int_{-\infty }^{\infty }dy\,\eta_{n}^{\ast }H_{1}\eta _{n} \,.\label{eq:Qvarespilon}
\end{eqnarray}
Combining (\ref{eq:H+H1}) with (\ref{eq:QvarepsOvareps})-(\ref{eq:Qvarespilon}) leads to
\begin{eqnarray}
k_n(\omega)= \sqrt{\varepsilon_n^{(0)}+\delta \varepsilon_n } \label{eq:kQomega}
\end{eqnarray}
and the subindex $n$ in $k_n$ was introduced to indicate that the dispersion relations are going to be state dependent. Furthermore the corresponding group velocities can be calculated as
\begin{eqnarray}
v_n^{g} = \frac{1}{\frac{dk_n}{d\omega}} \,. \label{eq:vQ}
\end{eqnarray}
We present here three states explicitly.

{\em Ground state}. 

\begin{figure}
 \center{
 {\includegraphics[height=40mm]{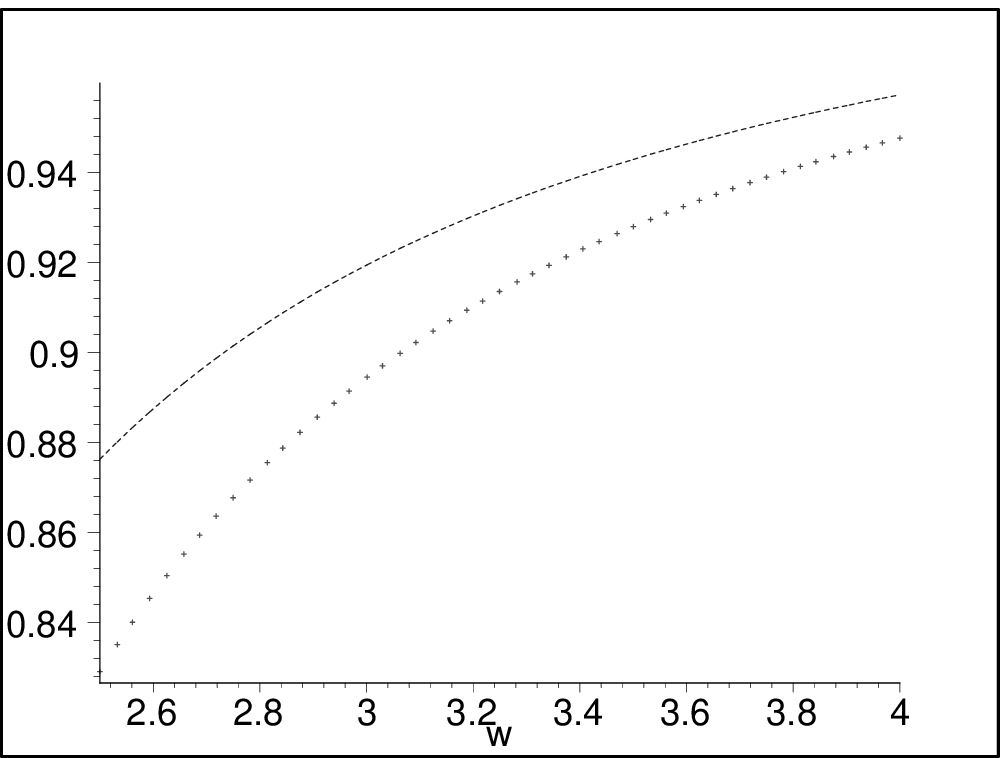}}
 \includegraphics[height=40mm]{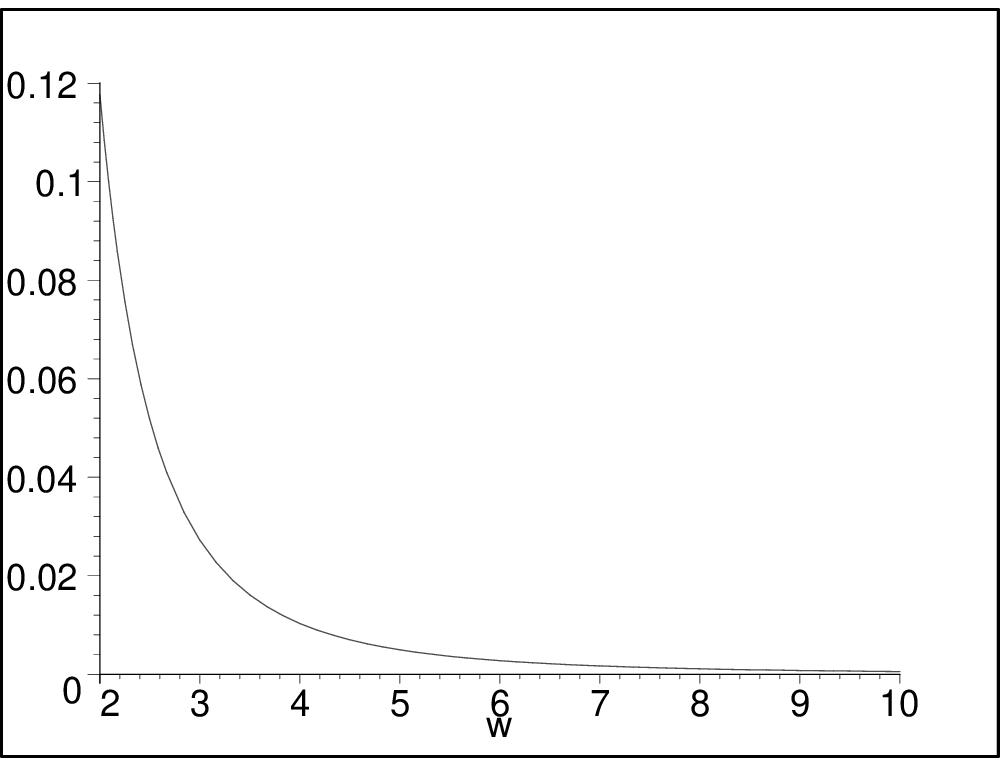}  } 
 \caption{{\small Ground state. Left: Group velocities of a charged particle (continuous curve) and a neutral particle (dashed curve). Right: $\frac{\delta\varepsilon_0}{\varepsilon_0}$. We take $E=1$, $M=1$, $e=1$. }}
\label{chi0}
\end{figure}

For this state the normalized eigenfunction is
\begin{eqnarray}
\eta _{0}&=&C_{0}\left( \cosh Ey\right) ^{-2\sigma } \\
(C_{0})^{-2}&=&\frac{1}{E}\frac{\Gamma (\frac{\omega }{E}-\frac{1}{2})\Gamma
(\frac{1}{2})}{\Gamma (\frac{\omega }{E})} 
\end{eqnarray}
Applying (\ref{eq:Qvarespilon}) to the ground state yields
\begin{eqnarray}
\delta \varepsilon_0 &=&
-\frac{e^2}{2}\frac{\sqrt{\pi}}{2\frac{\omega}{E}-1}\frac{\Gamma(\frac{\omega}{E}+
\frac{1}{2})}{\Gamma(\frac{\omega}{E}+1)}
\label{eq:deleps0}
\end{eqnarray}
with $\Gamma$ the gamma function. Fig.\ref{chi0} gives the comparison of the correspondent velocities of the particles, neutral and charged, in the ground state. Also the ratio $\frac{\delta\varepsilon_0}{\varepsilon_0}$ is shown to validate the perturbation theory.

\begin{figure}
\center{
{\includegraphics[height=40mm]{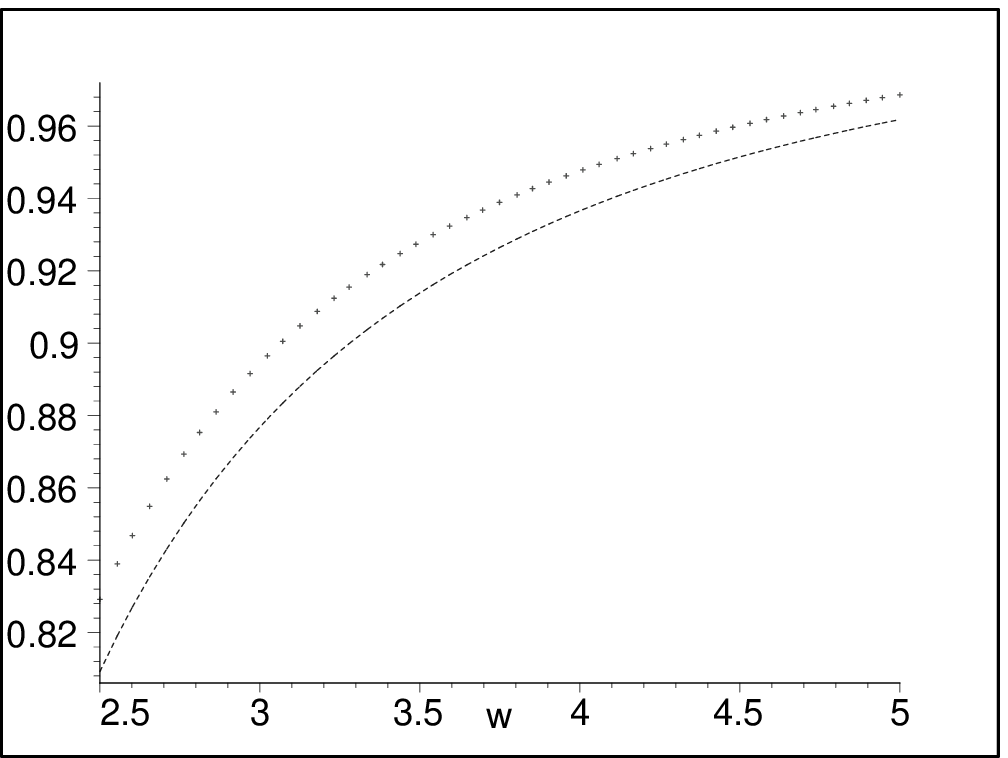}}
\includegraphics[height=40mm]{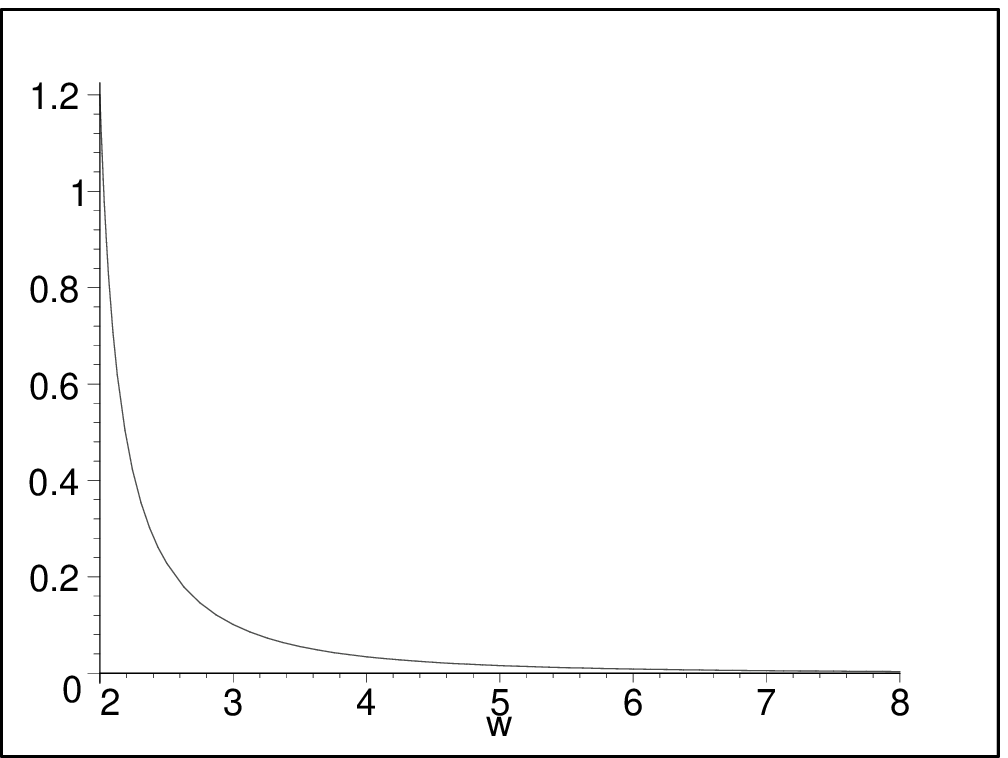}} 
\caption{
\small First excited state. Left: The dashed curve gives the group velocity of the neutral particle whereas the continuous curve is associated with the charged particle. Right: $\frac{\delta\varepsilon_1}{\varepsilon_1}$. We use $E=1$, $M=1$, $e=1$.} 
\label{chi1}
\end{figure}

{\em First excited state}.\\
In this case the wave function is 
\begin{eqnarray}
\eta _{1} &=&iC_{1}\sinh (Ey)(\cosh (Ey))^{-2\sigma } \\
(C_{1})^{-2} &=&\frac{\sqrt{\pi }}{E}
\left(3-2\frac{\omega}{E}\right)^{-1}
\frac{\Gamma (\frac{\omega }{E}-\frac{1}{2})}{%
\Gamma (\frac{\omega }{E})}
\end{eqnarray}
The corresponding contribution to the dispersion relations is
\begin{eqnarray}
\delta \varepsilon _{1} &=&
-\frac{e^2}{2}\frac{3\sqrt{\pi}}{(2\frac{\omega}{E}+1)(2\frac{\omega}{E}-1)}
\frac{\Gamma (\frac{\omega}{E}+\frac{1}{2})}{%
\Gamma (\frac{\omega }{E}+1)} \,.
\label{eq:deleps1}
\end{eqnarray}
The results are contained in Fig.\ref{chi1}

{\em Third excited state}.\\
Now the state is given by
\begin{eqnarray}
\eta_3&=& iC_{3}senh(Ey)[\cosh (Ey)]^{-2\sigma } \nonumber \\
&& +\frac{iC_{3}}{2}(2\kappa -1)[senh(Ey)]^{2}[\cosh
(Ey)]^{-2\sigma -2\kappa -3} \nonumber \\
(C_3)^{-2}&=&\frac{\sqrt{\pi}}{(3-2\omega)E}\frac{\Gamma(\omega-\frac{1}{2})}{\Gamma(\frac{1}{2})}-\frac{3}{32}\frac{(2\kappa
-1)^{2}}{(2\kappa +1)(\kappa +1)}\frac{\Gamma (2\kappa
+3)}{2\kappa +\frac{7}{2}} \nonumber \\
\kappa &=& \frac{1}{2}\sqrt{(\omega -\frac{7}{2})-\frac{1}{2}}
\,\,.
\nonumber \\
\end{eqnarray}
Now the correction to first order to the dispersion relations becomes
\begin{eqnarray}
\delta \varepsilon_3(\omega) &
=&-\frac{3e^{2}}{2}\frac{\sqrt{\pi }}{(3-2\omega )(2\omega
-1)}\frac{\Gamma (\omega +\frac{1}{2})}{\Gamma (\omega +1)}
\nonumber \\
&&-\frac{15e^{2}}{8}(2\kappa -1)^{2}\frac{\sqrt{\pi }}{(4\kappa
+2)(4\kappa +4)(4\kappa +6)}\frac{\Gamma (2\kappa +4)}{\Gamma
(2\kappa +\frac{9}{2})}\nonumber \\
&&3\sqrt{\pi }e\omega \frac{(2\kappa -1)}{(\omega +2\kappa
+\frac{1}{2})(\omega +2\kappa +\frac{5}{2})}\frac{\Gamma
(\frac{\omega }{2}+\kappa
+\frac{9}{4})}{\Gamma (\omega +\kappa +11)} \,.
\label{eq:deleps3}
\end{eqnarray}

\begin{figure}
\center{
\includegraphics[height=40mm]{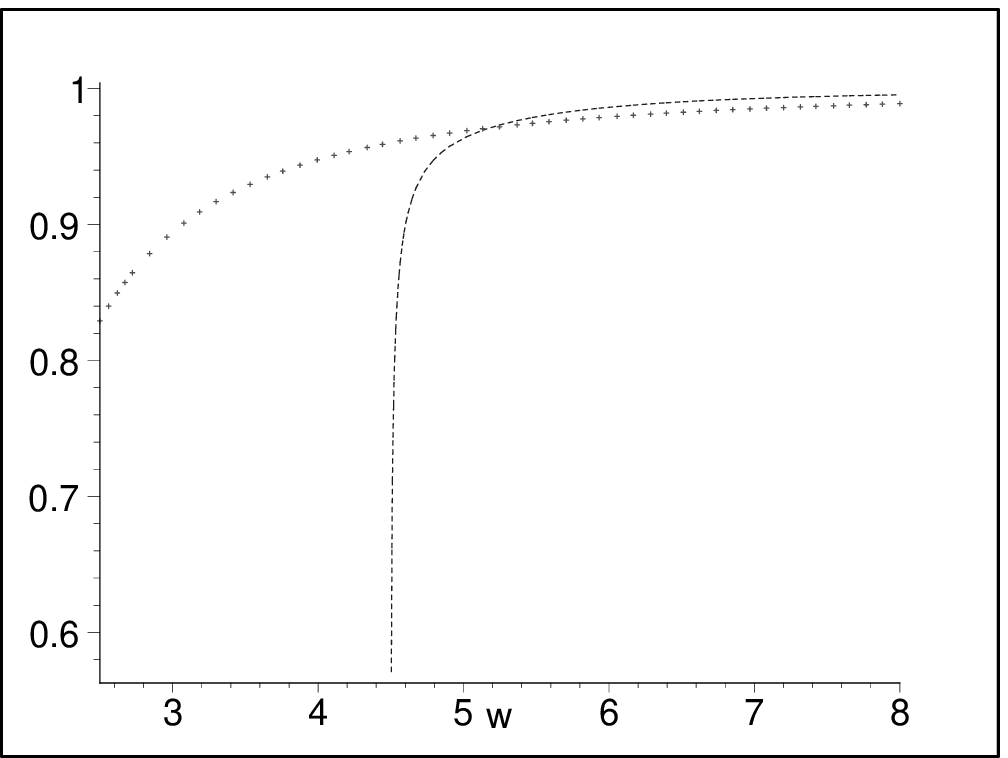}
\includegraphics[height=40mm]{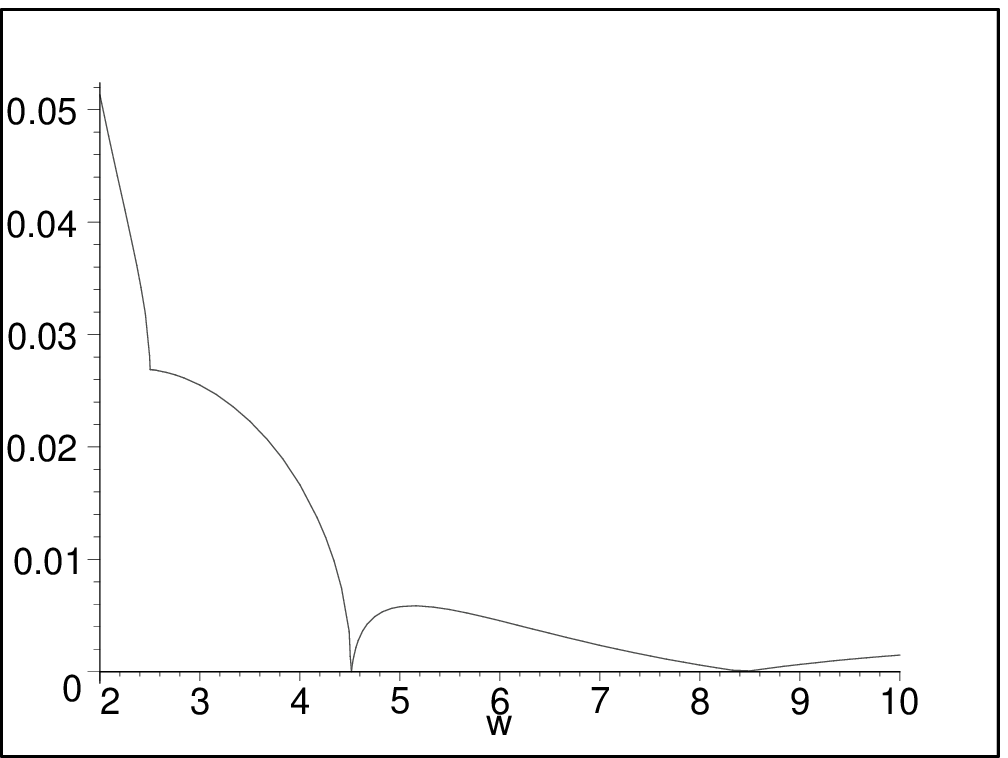}
}
\caption{\small 
Third excited state. Left: The dashed curve provides the velocity for the neutral particle while the continuous curve gives that of the charged particle. Right: $\frac{\delta\varepsilon_3}{\varepsilon_3}$.We use $E=1$, $M=1$, $e=1$.} 
\label{chi3}
\end{figure}

The comparison between the neutral and charged particles regarding group velocity appear in Fig.\ref{chi3}

\section{Discussion}
In this work we have studied the propagation of a charged scalar particle  propagating in the brane world model of Visser. This is an extension of \cite{visser} developed for a neutral scalar particle. Thus, in our case the particle is additionally coupled to the electric field pointing in the extra dimension that plays the role of source in the hyperdimensional scenario. 
 
 The dynamics of the charged particle given by the Klein-Gordon equation was effectively reduced to a Schr\"odinger type of equation. The associated potential, Fig.\ref{metaestados}, allow for bound states localized on the brane subspace located at $y=0$ but inhibits any tunneling into the extra dimension. It does not exclude though the situation corresponding to a particle propagating in the bulk that happens to cross the brane subspace at $y=0$. Lacking an exact solution of the correspondent eigenvalue problem we used a first order perturbative approximation that accounts for the charge of the particle in the effective potential with respect to the one of the neutral particle.

We have calculated the dispersion relations  and group velocities (Eqs. (\ref{eq:kQomega}), (\ref{eq:vQ}) and (\ref{eq:deleps0}), (\ref{eq:deleps1}), (\ref{eq:deleps3})) associated to the ground, first and third excited states. Although we could calculate them for any eigenstate we did not recognize any regular pattern to present them in a simple way. On the other hand the similarity of the results for the second excited state with those of the first led us to present here instead the third excited state. 

The complicated form of the first order corrected dispersion relations indicates a 4D Lorentz invariance violation, as opposed to the case of a neutral scalar particle propagating in the same scenario \cite{visser}. However, it is not clear whether a non perturbative analysis might change this result \cite{Dubovsky}.

 It is worth stressing that while the electric field in the Visser scenario points in the extra dimension it should affect a charged particle even though it moves constrained to four dimensions. Thus the charged particle we observe in four dimensions carry an imprint of the hyperdimensional scenario.  

It would be interesting to study the behavior of charged spin different from zero particles in Visser scenario and compare the results including the present ones with the current literature \cite{Dubovsky}.

\section*{Acknowledgements}
This work has been supported by Grant CONACyT-32431-E. RG acknowledges the support of a CONACyT fellowship .

\end{document}